\documentclass[twocolumn,english,prc,nofootinbib,floatfix,showpacs]{revtex4}
\usepackage[T1]{fontenc}
\usepackage[latin9]{inputenc}
\usepackage{amsmath}
\usepackage{graphicx}
\usepackage{amssymb}
\usepackage{esint}

\makeatletter
\@ifundefined{textcolor}{}
{%
 \definecolor{BLACK}{gray}{0}
 \definecolor{WHITE}{gray}{1}
 \definecolor{RED}{rgb}{1,0,0}
 \definecolor{GREEN}{rgb}{0,1,0}
 \definecolor{BLUE}{rgb}{0,0,1}
 \definecolor{CYAN}{cmyk}{1,0,0,0}
 \definecolor{MAGENTA}{cmyk}{0,1,0,0}
 \definecolor{YELLOW}{cmyk}{0,0,1,0}
 }

\usepackage{graphics}\usepackage{epsfig}\usepackage{amsfonts}
\usepackage{bm}
\@ifundefined{definecolor}
 {\usepackage{color}}{}
\renewcommand{\vec}{\bm}
\newcommand{\pr}[1]{{\sc{\lowercase{#1}}}}

\makeatother

\usepackage{babel}

\begin{document}

\title{Fluctuating parts of nuclear ground state correlation energies }

\author{B.~G.~Carlsson$^{1}$, J.~Toivanen$^{2}$ and U. von Barth$^{1}$}

\affiliation{$^{1}$ Division of Mathematical Physics, LTH, Lund University, Post
Office Box 118, S-22100 Lund, Sweden}

\affiliation{$^{2}$ Department of Physics, University of Jyväskylä, P.O. Box
35 (YFL) FI-40014, Finland}

\email{gillis.carlsson@matfys.lth.se}

\date{\today}
\begin{abstract}
\textbf{Background:} Heavy atomic nuclei are often described using
the Hartree-Fock-Bogoliubov (HFB) method. In principle, this approach
takes into account Pauli effects and pairing correlations while other
correlation effects are mimicked through the use of effective density-dependent
interactions. 

\textbf{Purpose:} Investigate the influence of higher order correlation
effects on nuclear binding energies using Skyrme's effective interaction.

\textbf{Methods:} A cut-off in relative momenta is introduced in order
to remove ultraviolet divergences caused by the zero-range character
of the interaction. Corrections to binding energies are then calculated
using the quasiparticle-random-phase approximation (QRPA) and second
order many-body perturbation theory (MBPT2).\textbf{ }

\textbf{Result:} Contributions to the correlation energies are evaluated
for several isotopic chains and an attempt is made to disentangle
which parts give rise to fluctuations that may be difficult to incorporate
on the HFB level. The dependence of the results on the cut-off is
also investigated.\textbf{ }

\textbf{Conclusions:} The improved interaction allows explicit summations
of perturbation series which is useful for the description of some
nuclear observables. However, refits of the interaction parameters
are needed to obtain more quantitative results. 
\end{abstract}

\pacs{21.10.Dr, 21.30.Fe, 21.60.Jz }

\maketitle

\section{Introduction }

The atomic nucleus is a complicated quantum mechanical system where
the probability to find a nucleon in a certain position is a function
of the positions of the other nucleons. 
This is generally referred to as the nucleons being correlated and
makes the wave functions of heavy nuclei too complex to compute directly
using ab initio theory. One therefore has to resort to more tractable
methods which take the most important correlation effects explicitly
into account, i.e. the ones that are important in order to describe
observables, while the remaining effects are treated in an approximate
way. 

An often used starting point is to assume that the in-medium interaction
between nucleons can be modeled using effective density-dependent
internucleon potentials. Such potentials are generally employed in
Hartree-Fock-Bogoliubov (HFB) calculations and their parameters are
fitted to reproduce a number of experimentally known data on individual
nuclei and to what is known about nuclear matter. In this way, Pauli
effects and pairing type correlations are taken into account explicitly
while the effects of other of correlations are described in an average
way. This approach has the great advantage of being applicable to
the entire nuclear chart at a reasonable computational cost. In the
quest of more accurate nuclear models an important task however, is
the systematic investigation of which type of correlation effects
can be modeled with the HFB method and which need a more explicit
treatment.

Several studies have shown that going beyond a HFB treatment and adding
corrections to binding energies resulting from shape vibrations, especially
of quadrupole type, give an improved description of experiment \cite{Bender2006,Delaroche2010,Moller1981,Baroni2006}.
These corrections are often taken into account by either using the
generator-coordinate method (GCM) \cite{Bender2006} or through the
use of a collective model e.g. a Bohr Hamiltonian \cite{Delaroche2010}. 

Alternatively, many-body perturbation theory (MBPT) offers a way to
explicitly and pictorially include elementary processes that one might
suspect to be responsible for correlations in different systems. For
instance, within the so called random-phase approximation (RPA) one
allows for an infinite number of particle-hole pairs to be excited
out of the Hartree-Fock ground state and for multiple scattering between
excited particles and holes. If the excitations instead are made out
of the HFB ground state the same approximation is referred to as the
QRPA designating RPA for Bogoliubov quasi-particles. An even simpler
step beyond HFB is the second-order many-body perturbation theory
(MBPT2) starting from the HFB ground state. Clearly, the virtual excitations
included in this approximation is a subset of those included in the
QRPA and in the present work we show results from both levels of approximation. 


Most of the effective nucleon potentials involve contact terms, i.e,
interactions of zero range. This is certainly the case for interactions
of the Skyrme type and such interactions give rise to divergences
when going beyond the HF level. This can be seen e.g. by solving the
two-body problem for $^{2}H$ analytically using contact interactions.
Then the resulting binding energies become infinite \cite{Thomas1935,Geltman2011}.
The two methods used in this work for going beyond the HFB level include
infinite summations of intermediate states which inevitably leads
to the same divergences in connection with zero-range forces.

It is a major theme of the present work to eliminate such divergences
by introducing cut-offs in momenta for our chosen Skyrme-like interaction
potentials \cite{Carlsson_conf2011}. This procedure implicitly assumes
that structures in binding energies as functions of nucleon number
originate in correlation effects caused by the low-momentum part of
the internucleon forces. And according to the results of the present
work this assumption does not appear to be that far fetched.

This paper is organized as follows: in Sec. II the regularized Skyrme
interaction is introduced. In Sec III we discuss the treatment of
correlation effects using the Quasiparticle-random-phase approximation
(QRPA) and the MBPT2 method. In sects. IV and V we analyze and discuss
the results of our calculations.

\section{low-momentum interaction}

\subsection{Two-body interaction in the particle-hole channel}

A general two-body interaction that preserves the center of mass coordinate
of the interacting particles can be expressed as

\begin{eqnarray*}
\hat{V}\left(\vec{r}_{1}'\vec{r}_{2}'\vec{r}_{1}\vec{r}_{2}\right) & = & v\left(\vec{r}',\vec{r}\right)\delta\left(\vec{R}-\vec{R}'\right),\end{eqnarray*}
where $\vec{R}=\frac{1}{2}\left(\vec{r}_{1}+\vec{r}_{2}\right)$ and
$\vec{r}=\vec{r}_{1}-\vec{r}_{2}$ denotes the center of mass and
relative coordinates respectively. The part of the potential depending
on relative coordinates can be transformed to momentum space and for
this part we adopt Skyrme's expansion \cite{Skyrme1959} given by

\begin{eqnarray*}
 &  & \bar{v}\left(\vec{k}',\vec{k}\right)=\frac{1}{\left(2\pi\right)^{3}}\int e^{-i\vec{k}'\cdot\vec{r}'}v\left(\vec{r}',\vec{r}\right)e^{i\vec{k}\cdot\vec{r}}d\vec{r}'d\vec{r}\\
 & \simeq & \frac{1}{\left(2\pi\right)^{3}}\left[t_{0}\left(1+x_{0}P^{\sigma}\right)+\frac{1}{2}t_{1}\left(1+x_{1}P^{\sigma}\right)\left(k'^{2}+k{}^{2}\right)\right.\\
 & + & \left.t_{2}\left(1+x_{2}P^{\sigma}\right)\vec{k}'\cdot\vec{k}+iW_{0}\left(\vec{\sigma}_{1}+\vec{\sigma}_{2}\right)\cdot\vec{k}'\times\vec{k}\right].\end{eqnarray*}
In this expression we have omitted the tensor potential included in
Ref. \cite{Skyrme1959} since it is not used in the parametrizations
we will employ later. This expression can be viewed as the first terms
in a low-momentum expansion of the effective nuclear potential going
up to second order in relative momenta (compare e.g. \cite{Carlsson2008,Raimondi2011,Carlsson2010}
for higher order expansions). Although this form gives a reasonable
description of the low-momentum parts, the expansion becomes unrealistic
for large momentum transfers. As we will demonstrate later, for Hartree-Fock
calculations, only the low-momentum matrix elements are important
and the unphysical contributions generated by the expansion for higher
momenta can be ignored. 

For studies beyond the mean-field level however, the interaction gives
diverging results unless some kind of truncation is enforced e.g.
a truncation in excitation energy. Nevertheless, in some beyond mean
field calculations, such as QRPA calculations, the results for low-lying
states \cite{Carlsson2012} and giant resonances \cite{Vesely2012}
are in reasonable agreement with experiment indicating that the interaction
may indeed have a wider applicability beyond purely mean-field calculations%
\footnote{Note that in the calculations of low-energy excitations the discussed
divergences did not constitute a problem. In fact, it is mainly the
high-energy excitations which are modified by a momentum cut-off.%
}. 

In order to investigate how well higher order corrections can be described
using  the low-momentum part of Skyrme's interaction we follow Skyrme's
original suggestion \cite{Skyrme1959} and introduce a cut-off in
momenta. We replace his original interaction by

\begin{equation}
\bar{v}^{\left(\Lambda\right)}\left(\vec{k}',\vec{k}\right)=\bar{v}\left(\vec{k}',\vec{k}\right)\theta\left(\mbox{\ensuremath{\Lambda}}-k'\right)\theta\left(\mbox{\ensuremath{\Lambda}}-k\right),\label{eq:cutt-off}\end{equation}
which vanishes at momenta above $\Lambda$ (fm)$^{-1}$. In the limit
of a large $\Lambda$ one regains the results of the original untruncated
interaction, but for finite values, the cut-off regularizes the interaction
so that beyond-mean field calculations converge. The introduction
of the cut-off destroys the nice analytical properties of the zero-range
interaction and increases the computational cost of calculating matrix
elements.

\subsection{Two-body interaction in the particle-particle channel}

In the pairing channel we use the same finite-range separable-Gaussian
interaction as was used in our previous studies \cite{Carlsson2012,Vesely2012}.
Since this interaction has a finite range, no regularization is needed.
We adopt an isospin invariant form, active in the $T=1$ channel and
use the same range parameter ($a=0.66$ fm) as before. Since we will
only consider cases where neutrons are in open shells we tune the
pairing strengths to make the lowest neutron quasi-particle energies
to agree with the experimental gaps determined in \cite{Dobaczewski1995}.
The resulting isovector pairing strength becomes $560$ MeV(fm)$^{-1}$
when the SKX Skyrme parameters \cite{Brown1998} are used in the particle-hole
channel and somewhat larger ($640$ MeV(fm)$^{-1}$) when the SLy5
parameters \cite{Chabanat1998} are used.

\subsection{Density-dependent part of the particle-hole interaction}

Skyrme's expansion of the two-body potential is often used together
with a density-dependent zero-range potential which is intended to
describe missing three-body and higher order contributions as well
as giving a simple representation of missing many-body effects. The
density-dependent terms cause difficulties when going beyond the mean
field and different recipes to define a residual interaction exist
in the literature \cite{Duguet2003}. In this work we are mainly motivated
by the success of the Skyrme interaction in connection with RPA type
calculations and hence define the residual interaction as the so-called
RPA residual interaction using the second derivative of the HF energy
\cite{RingSchuck1980}

\begin{eqnarray*}
\tilde{v}_{pmqn} & = & \frac{\partial^{2}E_{HF}}{\partial\rho_{qp}\partial\rho_{nm}}\\
 & = & v_{pmqn}\left[\rho\right]\\
 & + & \sum_{jl}\rho_{lj}\left(\frac{\partial v_{mjnl}\left[\rho\right]}{\partial\rho_{qp}}+\frac{\partial v_{pjql}\left[\rho\right]}{\partial\rho_{nm}}\right)\\
 & + & \left.\frac{1}{2}\sum_{ijkl}\rho_{ki}\frac{\partial v_{ijkl}\left[\rho\right]}{\partial\rho_{nm}\partial\rho_{qp}}\rho_{lj}\right|_{\rho=\rho_{0}}.\end{eqnarray*}
The use of the RPA residual interaction in configuration interaction
type calculations has been thoroughly discussed and investigated before
using a non-regularized Skyrme interaction \cite{Waroquier1987}. 

The density-dependent two-body interaction is introduced in the standard
form \cite{Bender2003} \begin{eqnarray*}
\hat{V_{\rho}} & = & v_{\rho}^{\left(\Lambda\right)}\left(\vec{r}',\vec{r}\right)\left(\rho\left(\vec{R}\right)\right)^{\alpha}\delta\left(\vec{R}-\vec{R}'\right),\end{eqnarray*}
with a dependence on the nucleon density $\rho$ to some power $\alpha$
which take on different values for different parameterizations. The
part dependent on relative coordinates is expanded to lowest order
in relative momenta

\begin{eqnarray*}
\bar{v}_{\rho}^{\left(\Lambda\right)}\left(\vec{k}',\vec{k}\right) & = & \frac{1}{\left(2\pi\right)^{3}}\int e^{-i\vec{k}'\cdot\vec{r}'}v_{\rho}^{\left(\Lambda\right)}\left(\vec{r}',\vec{r}\right)e^{i\vec{k}\cdot\vec{r}}d\vec{r}'d\vec{r}\\
 & \simeq & \frac{1}{\left(2\pi\right)^{3}}\frac{t_{3}}{6}\left(1+x_{3}P^{\sigma}\right)\theta\left(\mbox{\ensuremath{\Lambda}}-k'\right)\theta\left(\mbox{\ensuremath{\Lambda}}-k\right),\end{eqnarray*}
and regularized with the same cut-off procedure ($\Lambda$-truncation)
as used for the density-independent parts. 

In the practical calculations of matrix elements we start from a spherical
Harmonic-oscillator basis and transform the basis functions to momentum
space. The $\Lambda$-truncation can then be implemented using the
Moshinsky transformation \cite{Kamuntavicius2001} to transform the
coupled two-particle states to functions of relative and total momenta.
Finally we employ the Pandya transformation \cite{Suhonen2007} to
obtain matrix elements in the particle-hole channel and use the Wigner-Eckart
theorem to obtain angular-momentum reduced expressions. The full implementation
of this new regularized potential was done by extending the program
\textsf{\pr{HOSPHE}} (v1.02) \cite{Carlsson2010p2}.

 {

One of the interactions employed in this work (SLy5) uses the direct
part of the Coulomb interaction together with a Slater approximation
for the Coulomb exchange. The Slater approximation results in a density-dependent
term which mimics the HF exchange energy. In order to treat the HF
part and the additional residual interaction consistently we have
regularized the Slater term in the same way as for the other parts
of the interaction. 

\begin{figure}[t]
\includegraphics[clip,width=1\columnwidth]{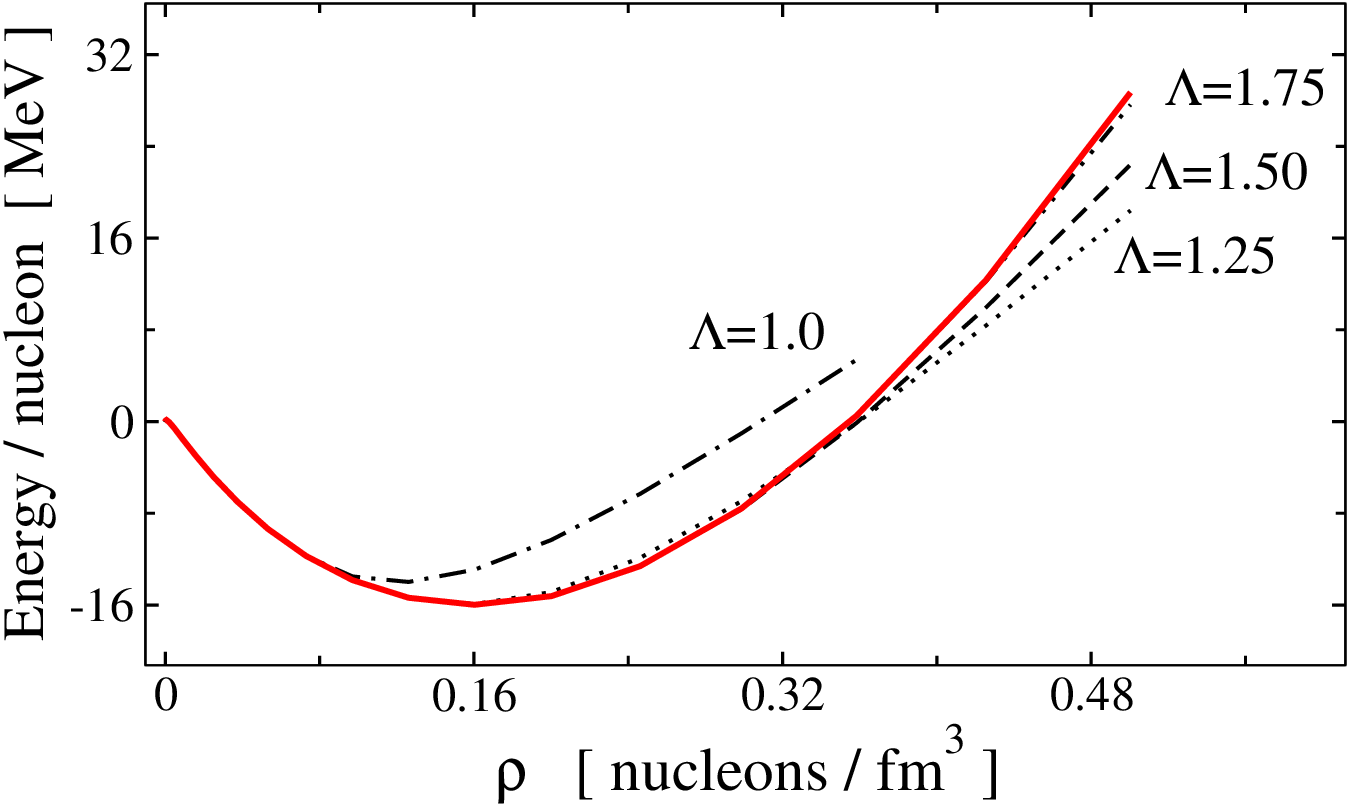}

\caption{(Color online) The energy per nucleon in symmetric nuclear matter
(equal numbers of neutrons and protons) shown for the SLy5 \cite{Chabanat1998}
interaction and for different $\Lambda$ (fm)$^{-1}$ values in the
Hartree-Fock approximation. The full line illustrates the standard
result for the untruncated interaction ($\Lambda=\infty$). \label{fig:EOS}}
\end{figure}
In order to have a first idea about the influence of the $\Lambda$
-truncation we consider isospin-symmetric nuclear matter in the Hartree-Fock
approximation. The corresponding zero-temperature equation of state
(EOS) i.e. the energy per nucleon as a function of density is shown
in Fig.~\ref{fig:EOS}. As seen in this figure, a $\Lambda$ value
of $\approx$1.5 (fm)$^{-1}$ leaves the EOS unchanged up to about
twice the saturation density and a value of $\approx$1.75 (fm)$^{-1}$
leaves the EOS unchanged up to about three times the saturation density.
Thus, when choosing a value for the regularization we will consider
values above $1.5$ (fm)$^{-1}$ which keeps the relevant part of
the EOS approximately the same.

\section{Results for the total correlation contribution}

An expression for the RPA correlation energy in the quasi-boson approximation
(QBA) was derived in \cite{RingSchuck1980} and using an analogous
derivation one obtains a corresponding expression in the QRPA case
\cite{Egido1980}

\begin{equation}
E_{QRPA}=-\sum_{\nu}\hbar\omega_{\nu}\sum_{k<k'}\left|Y_{kk'}^{\nu}\right|^{2}.\label{eq:QRPA1}\end{equation}
In order to evaluate this expression we start by defining matrices
containing the positive energy QRPA vectors\[
X=\left[X^{1},X^{2},...,X^{N}\right],\, Y=\left[Y^{1},Y^{2},...,Y^{N}\right]\]
and corresponding energies \[
\Omega=\left[\begin{array}{ccc}
\hbar\omega_{1} & 0 & 0\\
0 & \ddots & 0\\
0 & 0 & \hbar\omega_{N}\end{array}\right].\]
Then starting from the QRPA equation \cite{Suhonen2007}

\begin{equation}
\left(\begin{array}{cc}
A & B\\
B^{*} & A^{*}\end{array}\right)\left(\begin{array}{c}
X\\
Y\end{array}\right)=\left(\begin{array}{c}
X\\
-Y\end{array}\right)\Omega\label{eq:3-3}\end{equation}
one can write the pair of equations 

\begin{eqnarray*}
YX^{-1}A+YX^{-1}BYX^{-1} & = & Y\Omega X^{-1}\\
B^{*}+A^{*}YX^{-1} & = & -Y\Omega X^{-1}.\end{eqnarray*}
Summing these equations together and introducing $C=YX^{-1}$ the
result is the equation\begin{equation}
B^{*}+A^{*}C+CA+CBC=0\label{eq:QRPA}\end{equation}
which is similar to the multiple scattering series derived in Ref.
\cite{Mavromatis1991}. In terms of $C$, the QRPA correlation energy
becomes 

\begin{eqnarray}
E_{QRPA} & = & \frac{1}{2}\sum_{k<k',l<l'}B_{kk',ll'}C_{kk',ll'}.\label{eq:QRPACcorr}\end{eqnarray}
Furthermore, by splitting the $A$ matrix 

\[
A_{kk',ll'}=\left(E_{k}+E_{k'}\right)\delta_{kl}\delta_{k'l'}+\bar{A}_{kk',ll'},\]
Eq. \ref{eq:QRPA} can be written \begin{align}
C_{kk',ll'} & =\frac{-1}{E_{k}+E_{k'}+E_{l}+E_{l'}}\nonumber \\
 & \times\left(B_{kk',ll'}^{*}+\left(\bar{A}^{*}C\right)_{kk',ll'}+\left(C\bar{A}\right)_{kk',ll'}\right.\nonumber \\
 & \left.+\left(CBC\right)_{kk',ll'}\right).\label{eq:Ceq}\end{align}

Finally, assuming that an iteration procedure for $C$ converges,
we can evaluate $C$ order by order where the first order contribution
\[
C_{kk',ll'}^{\left(1\right)}=\frac{-B_{kk',ll'}^{*}}{E_{k}+E_{k'}+E_{l}+E_{l'}}\]
is obtained by putting $C$ equal to zero on the right hand side of
Eq. \ref{eq:Ceq}. Higher orders $C^{\left(n\right)}$ are thus obtained
by repeatedly inserting the previous expression $C^{\left(n-1\right)}$
on the right hand side. By using $C^{\left(n\right)}$ in the formula
for the correlation energy, Eq. \ref{eq:QRPACcorr}, we obtain $E_{QRPA}^{\left(n\right)}$.
Numerically we have also verified that the iteration converges to
results consistent with Eq. \ref{eq:QRPA1}. 

An alternative approach which is less costly numerically is to evaluate
the correlation contribution from second order perturbation theory
\cite{Sakurai1994} starting from the HFB ground state and treating
the residual part of the quasi-particle Hamiltonian as a perturbation.
In this case, the only contributions that arise come from scattering
to four quasi-particle states via the $H^{40}$ part \cite{RingSchuck1980}
of the Hamiltonian. This contribution can be expressed in terms of
the QRPA $B$ matrix according to \begin{align}
E_{MBPT}^{\left(2\right)} & =-\frac{1}{6}\sum_{k<k',l<l'}\frac{\left|B_{ll'kk'}\right|^{2}}{E_{k}+E_{k'}+E_{l'}+E_{l'}}.\label{eq:MBPT2}\end{align}

It is interesting to compare the QRPA and MBPT series order by order.
The lowest order QRPA term is three times larger than $E_{MBPT}^{\left(2\right)}$
while the third order $E_{MBPT}^{\left(3\right)}$ is exactly obtained
in the QRPA series. In higher orders, the two series differ as the
QRPA expression only includes a subsequence of the full MBPT series. 

In earlier papers by Ellis \cite{Ellis1970,Ellis1987}, the RPA correlation
energy was investigated by starting from an unpaired ground state
and summing contributions from both normal and pairing vibrations
using diagrammatic techniques. In this way it was shown that in the
QBA, the second order contribution appears twice in the summation
of the particle-hole ring series and once in the particle-particle
series. A suggested remedy for this overcounting was to remove the
second-order term from the particle-hole series and only keep it in
the particle-particle series. In this work we have however refrained
from using this approach since it is not directly applicable when
starting from a HFB state where normal and pairing vibrations are
generally mixed. 

\begin{figure}[t]
\includegraphics[clip,width=1\columnwidth]{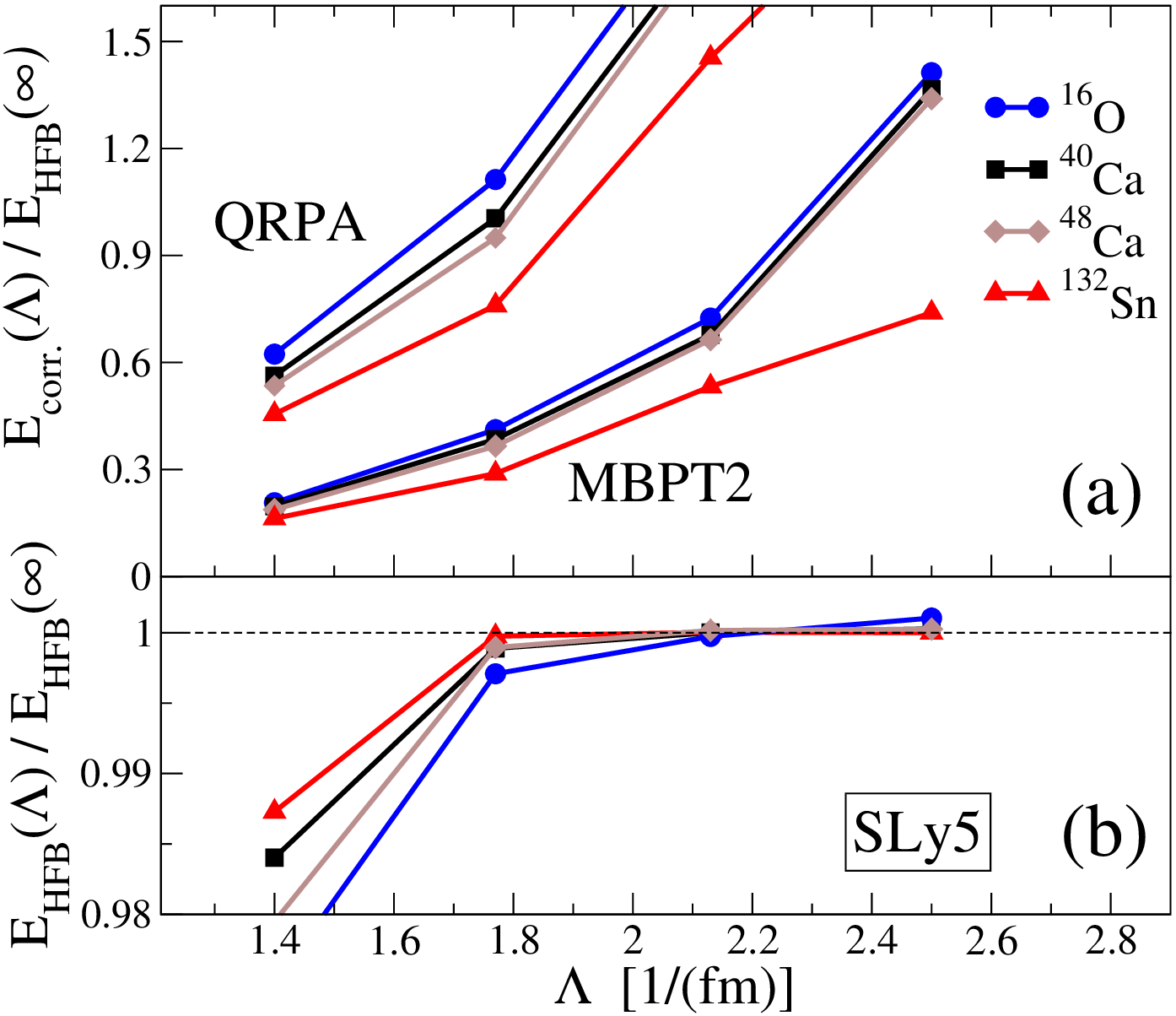}

\caption{(Color online) The upper panel shows the binding energy contribution
from QRPA and MBPT2 scaled with the HFB energy. The lower panel shows
the influence of the truncation on the HFB energy. \label{fig:totcorr}
Calculations were done using the SLy5 \cite{Chabanat1998} parametrization
for the Skyrme interaction using a basis of 14 oscillator shells. }
\end{figure}

The corresponding correlation energies evaluated with the two methods
described above are shown in panel (a) of Fig. \ref{fig:totcorr}.
In both results we have not included the part of the $B$ matrix obtained
in pnQRPA \cite{Suhonen2007} which is associated with excitations
of proton-neutron pairs. Although this contribution is certainly interesting,
a first step in the direction of including these effects would involve
tuning the effective interactions in the $T=0$ pairing channel. 

As seen in Fig. \ref{fig:totcorr}, for both methods the correlation
energy amounts to a rather large part of the total binding energy.
The QRPA formula predicts the largest values as the QBA overestimates
the ground state correlations \cite{BlaizotRipka1986,RingSchuck1980,Ellis1987,Ellis1970}.
Although this could possibly be corrected for, in the following we
will instead focus on the MBPT2 results.

In the case of $^{16}$O, the smallest $\Lambda=1.4$ (fm)$^{-1}$
used in the figure gives a contribution from MBPT2 which is 21 \%
of the HFB energy. This contribution gradually decreases for the heavier
nuclei and becomes 16 \% in $^{132}$Sn. 

Panel (b) of Fig. \ref{fig:totcorr} shows the influence the regularization
has on the HFB energy. As seen in this figure, the HFB energy converges
to the untruncated value as the cut-off is increased and even for
rather low cutoffs of $\Lambda=1.5$ (fm)$^{-1}$ the change in total
binding energy stays within a few percent. This tells us that the
HFB energy is not very sensitive to higher momentum parts of the potential
in the particle-hole channel. 

In both the QRPA and in MBPT2, the correlation energy increases rapidly
with increasing $\Lambda$ and in the following we will consider $\Lambda$
values in the range of 1.6-1.8 (fm)$^{-1}$. These values lead to
the smallest correlation energies while causing moderate changes of
the HFB energies.

Since the HFB energy stays roughly the same, it is a good approximation
to neglect the regularization for the HFB part of the calculation
and only regularize when generating the residual interaction. This
is quite important in order for the method to be practical since otherwise
one would have to recalculate the regularized density-dependent interaction
in each HFB iteration. This is a strong motivation for introducing
the regularization in the way done here rather than using for example
a Gaussian interaction. For the purpose of making the least amount
of approximations, we have however used the time-consuming strategy
to recalculate the regularization in each HFB step. 

Using an angular-momentum coupled notation, the correlation energy
can be divided into partial contributions arising from QRPA excitations
with different total angular momentum $J$ and parity $\pi$. Since
the MBPT2 result can be seen as an approximation to the full QRPA
results we use the same division into multipole contributions also
in this case. These partial contribution are shown in Fig.~\ref{fig: Ecorr_Vs_J}
for $^{132}$Sn. As seen in this figure the largest contributions
come from natural parity states with $\left(-1\right)^{J}=\pi$. Both
positive parity and negative parity contributions are equally important
and show maxima for $J=4$ and $J=5$ respectively. %
\begin{figure}[t]
\includegraphics[clip,width=1\columnwidth]{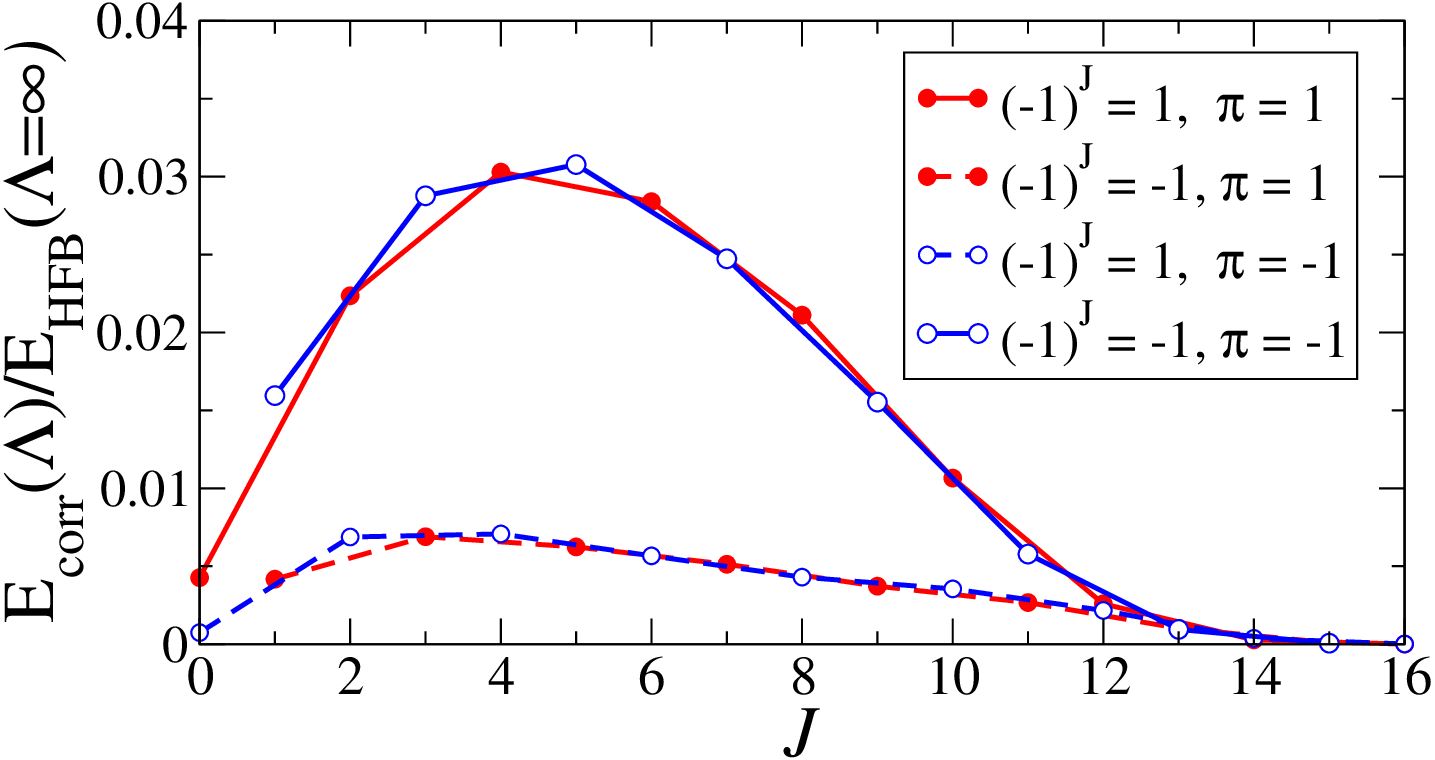}

\caption{(Color online) Partial contributions to the MBPT2 correlation energy
for $\Lambda=1.8$ and $^{132}$Sn using SLy5. \label{fig: Ecorr_Vs_J}}
\end{figure}

\section{Fluctuating parts of the correlation energy}

The internucleon potentials we employ have parameters which are fitted
in order to give reasonable nuclear properties at the HFB level. Therefore
the average part of the correlation energies is already effectively
included through the fitting of the model parameters. The total ground-state
energy can be divided into a liquid-drop part that captures the average
variations as a function of nucleon numbers and a fluctuating part
that mainly depends on the shell structure. Rather than performing
a full refit of the interaction to have a model on the MBPT2 level,
in this first study we will make a simple compensation for this overbinding.
We compensate by fitting a liquid-drop expression \cite{RingSchuck1980}
\begin{equation}
E_{LD}=a_{vol}A+a_{surf}A^{2/3}+a_{sym}\frac{\left(N-Z\right)^{2}}{A}\label{eq:LD}\end{equation}
to the correlation energy $E_{MBPT}^{\left(2\right)}$ which is then
subtracted to give the fluctuating part of the correlation energy
$\Delta E=E_{MBPT}^{\left(2\right)}-E_{LD}$. The main goal is to
get an idea what kind of fluctuations one obtains and to see if these
are correlated with errors obtained in the description of ground-state
energies.

\begin{figure}[t]
\includegraphics[bb=0bp 0bp 702bp 522bp,clip,width=1\columnwidth]{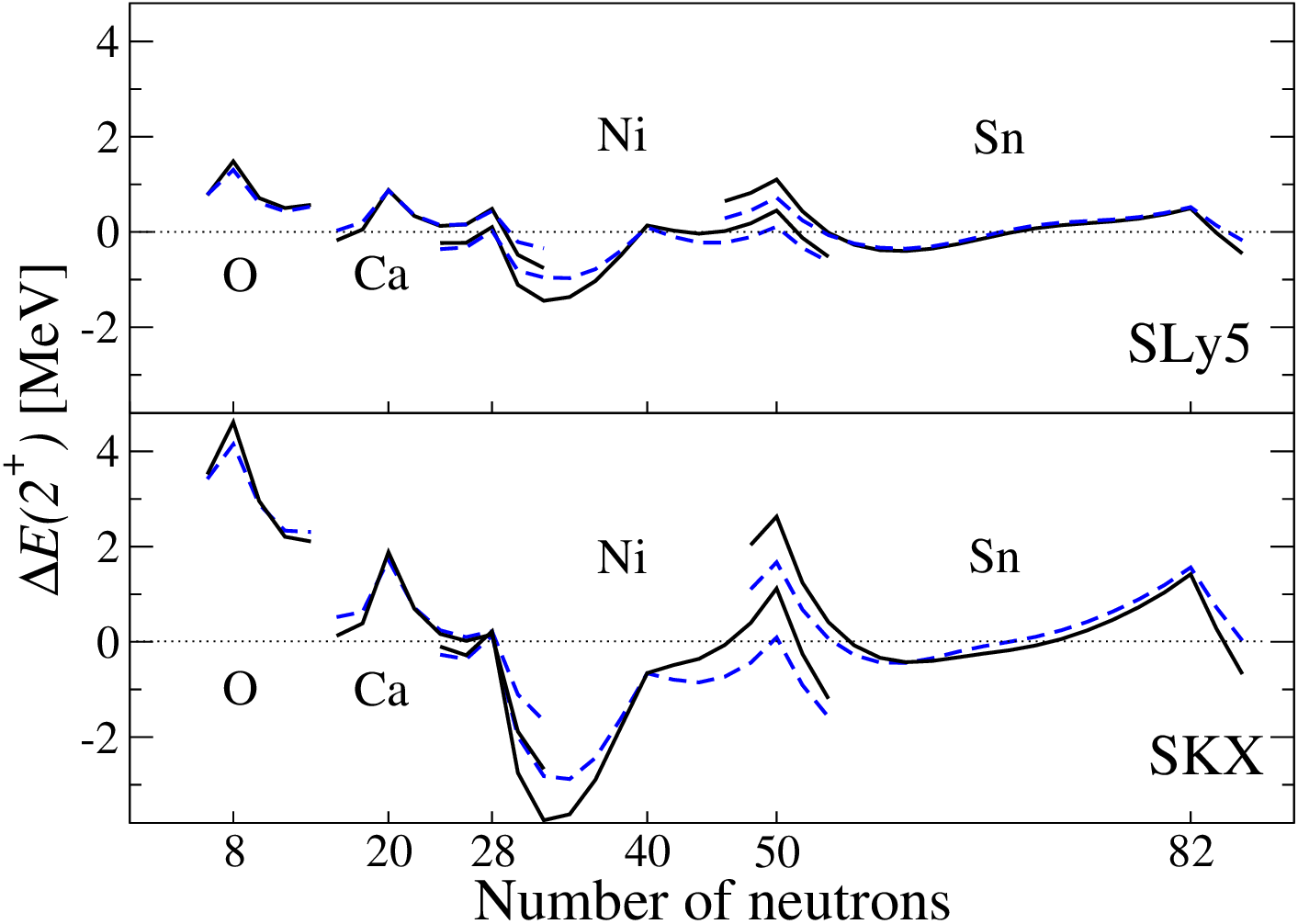}

\caption{(Color online) The renormalized part of the MBPT2 correlation energy
associated with quadrupole shape vibrations ($J=2^{+}$) shown for
$\Lambda=1.6$ (dashed curve) and $\Lambda=1.8$ (full curve). \label{fig: Ecorr_2p}}
\end{figure}
\begin{figure}[t]
\includegraphics[width=1\columnwidth]{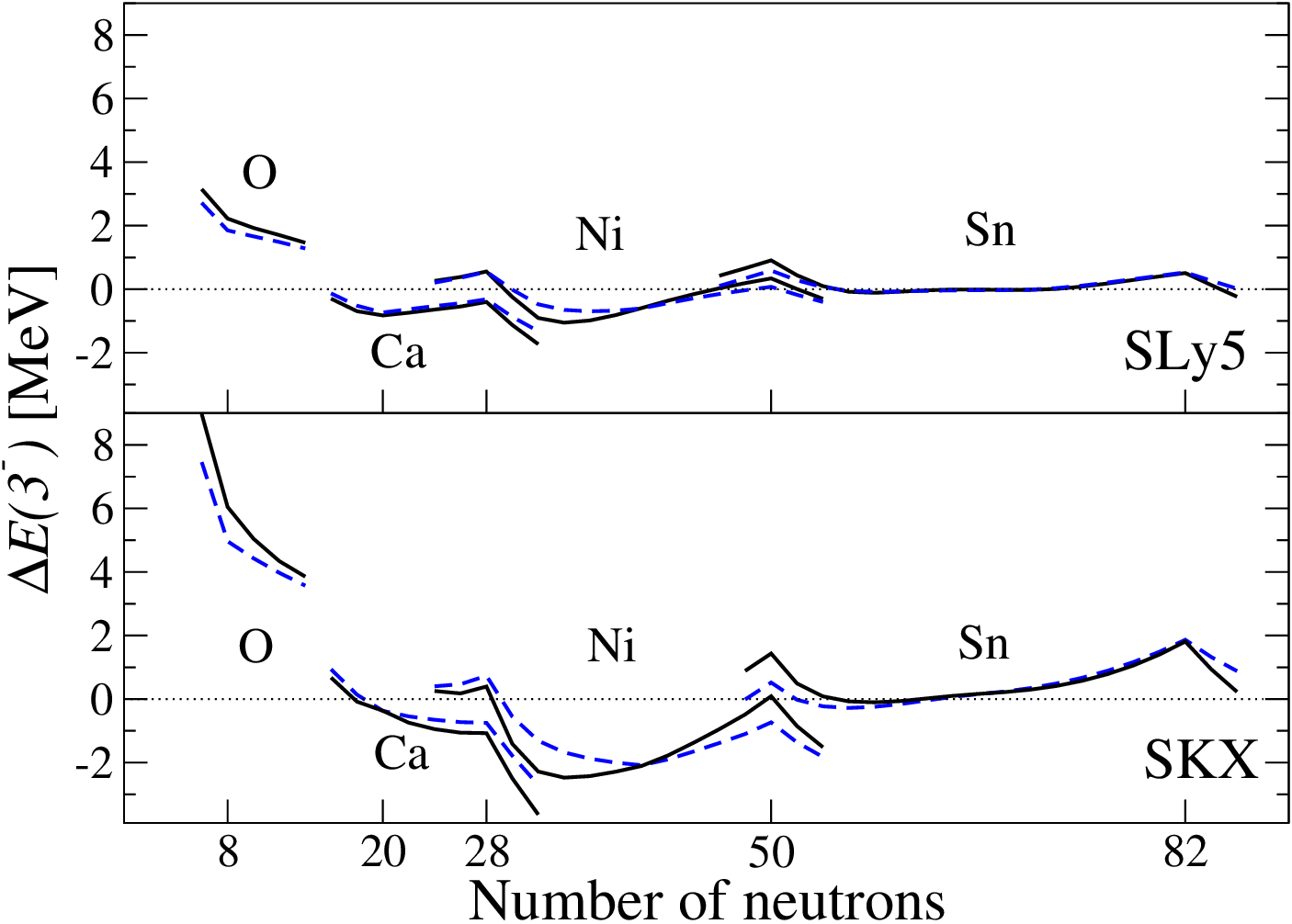}

\caption{(Color online) Same as Fig. \ref{fig: Ecorr_2p} for octupole shape
vibrations ($J=3^{-}$).\label{fig:Ecorr_3m}}
\end{figure}
In this way the renormalized correlation contributions associated
with $2^{+}$ and $3^{-}$ vibrations are extracted and shown in Figs.
\ref{fig: Ecorr_2p} and \ref{fig:Ecorr_3m}. The fluctuations show
pronounced shell effects and tend to give increased binding energy
contributions for the open shell nuclei compared to the magic ones.
The two different choices of $\Lambda$-truncation shown in the figures
give similar results indicating that the obtained fluctuations are
mainly associated with the properties of the low-momentum part of
the interaction.

In the $2^{+}$ channel, the SKX interaction gives larger fluctuations
than the SLy5 interaction. Previous results for quadrupole correlations
using the SkI3 interaction and a collective Hamiltonian \cite{Klupfel2008}
gave similar results with fluctuations that are somewhere in between
the ones we get for the SKX and the SLy5 interactions. 

Contributions from octupole vibrations are similar in magnitude to
the quadrupole vibrations and show the same tendency of increasing
the energies for magic nuclei as compared to their neighbors. Notable
exceptions are $^{16}$O and $^{40}$Ca which have Fermi levels between
opposite parity shells and show the reversed trend. 

\begin{figure}[t]
\includegraphics[clip,width=1\columnwidth]{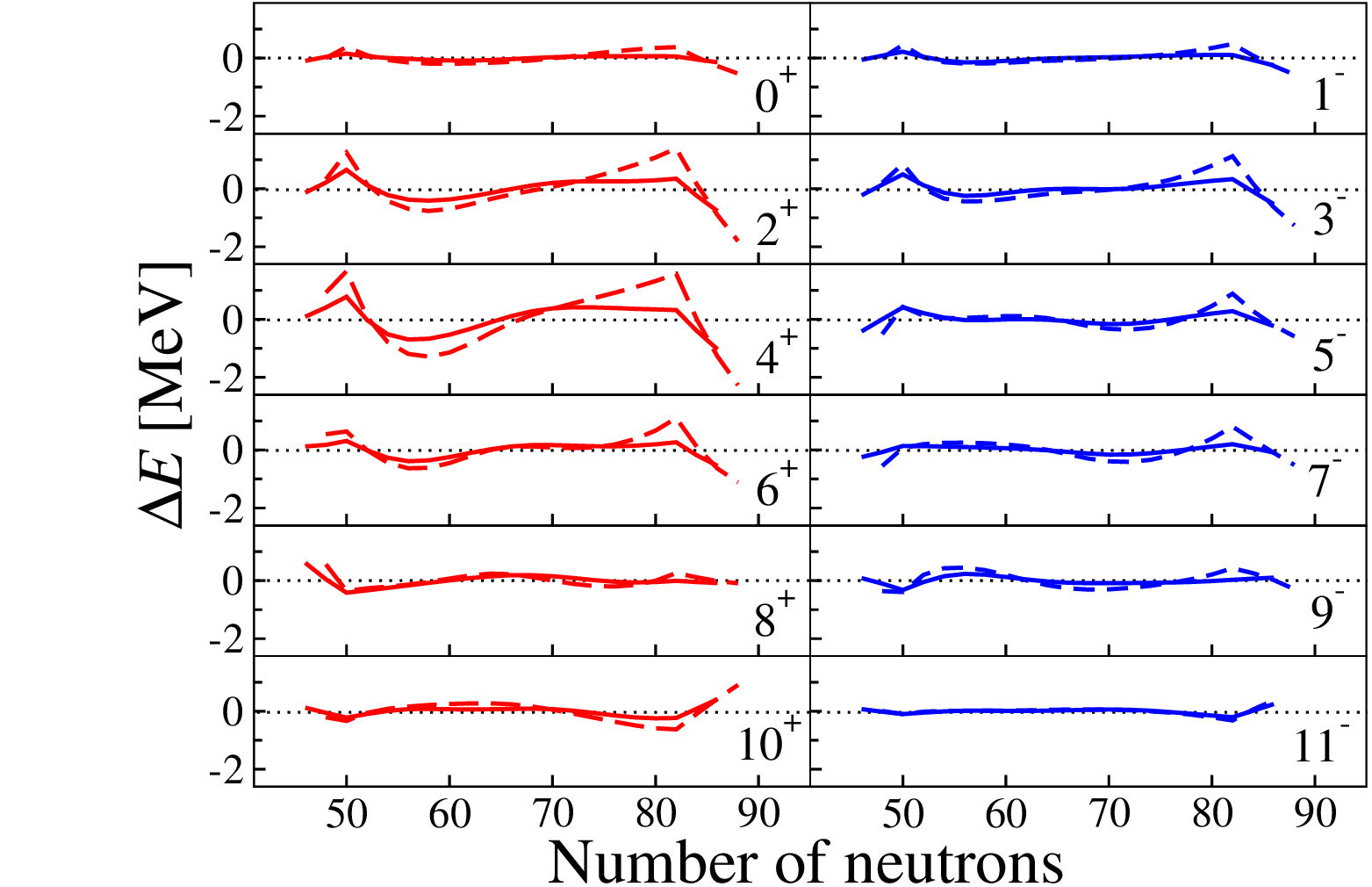}

\caption{(Color online) The renormalized part of the MBPT2 correlation energy
separated into contributions from different multipoles for $\Lambda=1.8$.
Solid curves are for SLy5 and dashed curves for SKX. In this figure
the renormalization was done by fitting Eq. \ref{eq:LD} to Sn nuclei
only. \label{fig: de_Sn}}
\end{figure}

Fig. \ref{fig: de_Sn} shows the fluctuating part of the correlation
energies separated into contributions from different multipoles in
the case of the Sn chain. The main fluctuations come from the $J^{\pi}=2^{+},4^{+},3^{-}$
and $5^{-}$channels and show the trend of making double-magic nuclei
less bound relative to the semi-magic ones. Interesting exceptions
are found in the $10^{+}$ and $11^{-}$ contributions which show
the opposite trend around $N$=82. Going even higher in multipoles
the curves tend to flatten out.

\subsection{Comparison with experiment}

\begin{figure}[t]
\includegraphics[clip,width=1\columnwidth]{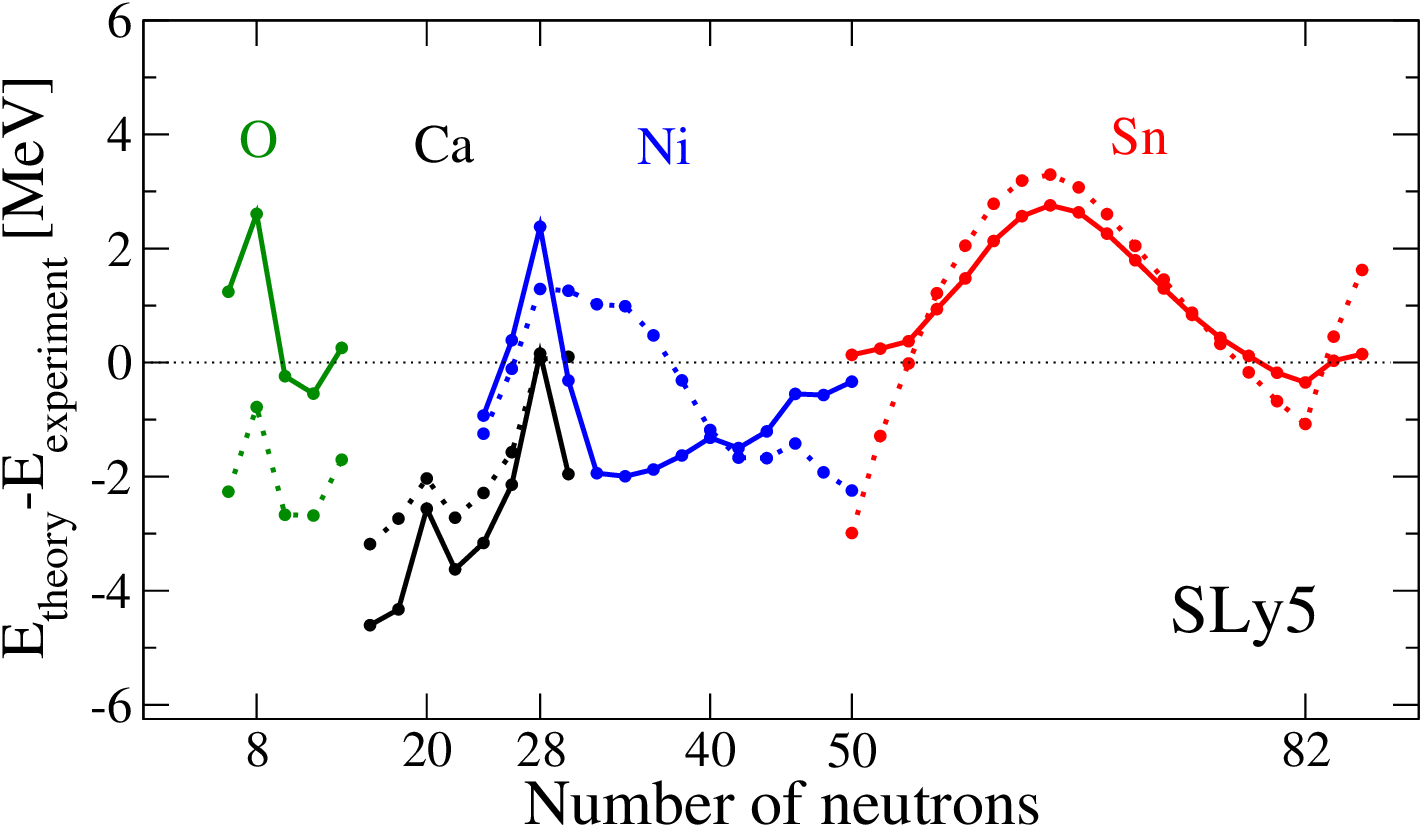}

\caption{(Color online) Difference between experimental \cite{Audi2003,ENSDF}
and theoretical ground-state energies in the HFB approach (dotted
curves) and when adding surface vibrations corresponding to multipolarities
$J^{\pi}=0^{+},2^{+},1^{-}$ and $3^{-}$ (full curves). Figure is
drawn using the SLy5 interaction and $\Lambda=1.8$ (fm)$^{-1}$.\label{fig: experiment1}}
\end{figure}
\begin{figure}[t]
\includegraphics[clip,width=1\columnwidth]{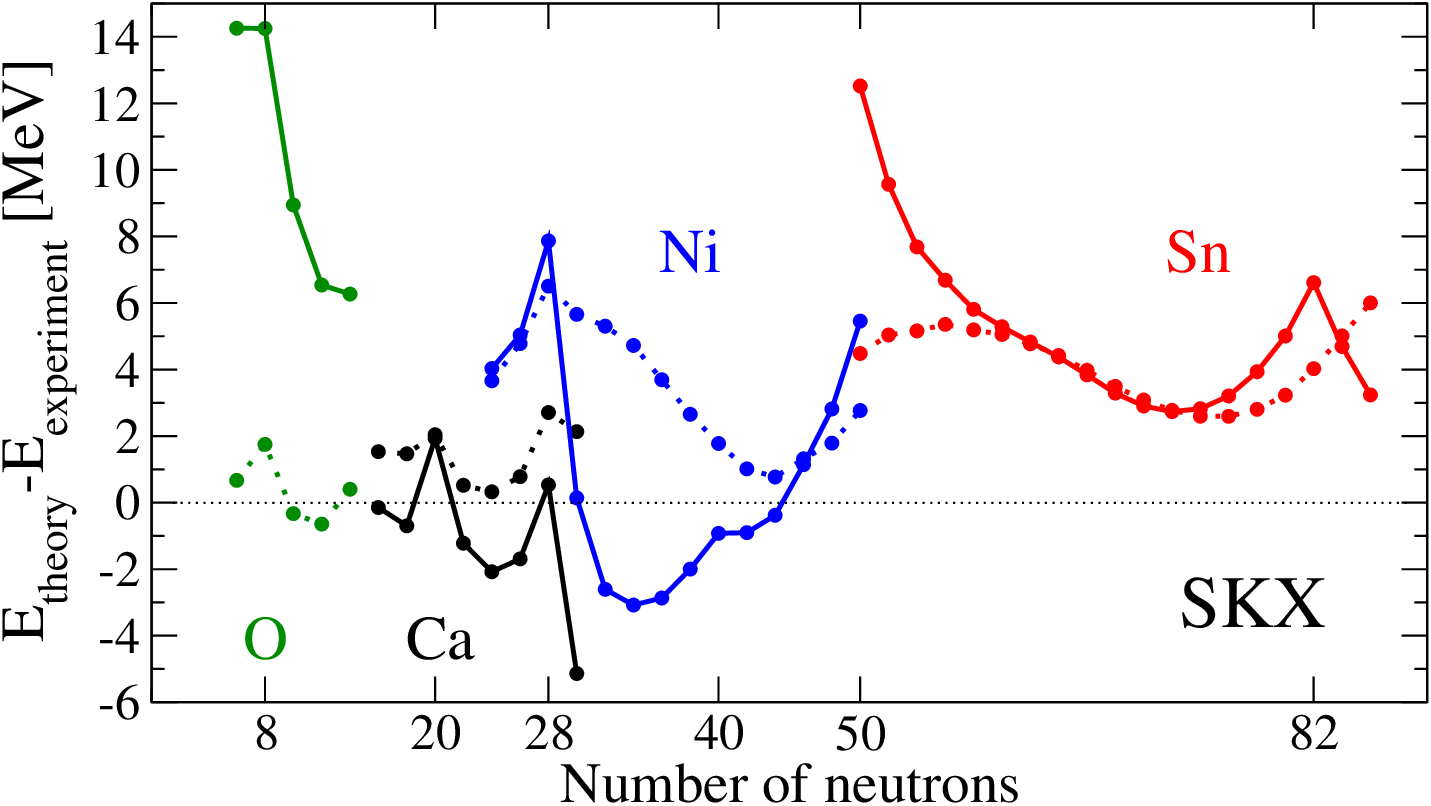}

\caption{(Color online) Same as \ref{fig: experiment1} but for the SKX interaction.\label{fig: experiment1-1}}
\end{figure}


When comparing with experimental ground-state energies, we restrict
ourselves to contributions from the well studied low-order multipoles
$J^{\pi}=0^{+},2^{+},1^{-}$ and $3^{-}$ where the effective interactions
generally give reasonable results for low-lying collective states
and giant resonances. 

The difference between calculated and experimental ground-state energies
using the SLy5 and the SKX \cite{Brown1998} interactions are shown
in Figs. \ref{fig: experiment1} and \ref{fig: experiment1-1}. The
dotted lines denote the results of the HFB treatment ($\Lambda=\infty$)
using a total of 31 oscillator shells. For both interactions, the
HFB results are within a few MeV of the experimental values. The SKX
results differ somewhat from the ones in \cite{Brown1998} which is
due to a different treatment of the Coulomb interaction. While we
calculate the direct Coulomb contribution directly from the proton
density, in \cite{Brown1998} some additional corrections were taken
into account.

It interesting to notice that with both interactions the errors for
magic nuclei with $N=8,20$ and $28$ go up in energy compared to
their neighbors which could possibly be cured by making the corresponding
gaps in the neutron spectra somewhat larger. With SLy5, the situation
is reversed for the gaps at $N=50$ and $82$ where the errors instead
dip down. 

The lowest order surface vibrations are calculated using MBPT2 and
a total of 14 oscillator shells. The fluctuating part of the correlation
energy is well converged using 14 oscillator shells and is extracted
by fitting the liquid-drop expression to all the included nuclei.
The solid lines in Figs. \ref{fig: experiment1} and \ref{fig: experiment1-1}
show the results of adding these fluctuating parts to the HFB energies.
As discussed previously the main effect of the surface vibrations
is to push the magic nuclei up in energy as compared to the neighboring
nuclei. For SLy5 this gives corrections that go in the right direction
in the region of the $N=50$ and $82$ gaps but in the opposite direction
for the lighter magic nuclei. Although both interactions give rise
to similar fluctuations, the larger magnitude fluctuations in combination
with different HFB results obtained using the SKX parametrization,
compares less favorable with experiment. If a Skyrme interaction tuned
at the MBPT2 level is used on the HFB level, one would expect it to
predict the magic nuclei to be more bound than their neighbors in
order to leave room for the additional correlation part.

Some of the nuclei included in the plot have the same number of neutrons
and protons which gives rise to an additional contribution to the
binding energy. This contribution is often modeled by adding so called
Wigner corrections (see e.g. \cite{Goriely2002}) which gives an additional
binding energy contribution of roughly 2 MeV for the $N=Z$ nuclei.
Such a contribution would reduce some of the remaining fluctuations
but would not improve the results around $^{48}$Ca for example. Furthermore,
such a phenomenological treatment is clearly unsatisfactory and a
more thorough investigation of these interesting effects is clearly
called for. 


In the case of $\Lambda=1.8$ (fm)$^{-1}$ the parameters obtained
from the liquid-drop fit to the SLy5 correlation energies resulting
from $J=0^{+},2^{+},1^{-}$ and $3^{-}$ vibrations become $\left\{ a_{vol},a_{surf},a_{sym}\right\} =\left\{ 0.99,-8.24,0.77\right\} $
MeV, while for SKX the average contribution is roughly twice as large.
A possible reason that the SKX interaction gives more correlation
energy is that SKX has larger effective mass ($m*/m=0.99$) \cite{Brown1998}
than SLy5 ($m*/m=0.69$) \cite{Chabanat1998} and thus a denser spectrum,
giving smaller denominators in Eq. \ref{eq:MBPT2}. 

Typical liquid-drop parameters obtained when fitting to experimental
ground-state energies are $\left\{ a_{vol},a_{surf},a_{sym}\right\} =\left\{ -15.68,18.56,28.1\right\} $
MeV \cite{RingSchuck1980}. Thus, the main change in the average energy
obtained by adding the correlations resulting from low-lying surface
vibrations is to modify the surface energy. The reduction of the surface
energy and increased energy for the volume part will likely move nucleons
from the bulk to the surface leading to a more diffuse surface region.
Thus refitting the Skyrme parameters to absorb the average part of
the correlations and have a model on the MBPT2 level would likely
involve tuning not only the density-dependent terms but also the gradient
terms which are more sensitive to the surface region. When refitting,
it is important to have as small correlation corrections as possible,
so that the HFB ground state is a reasonable first approximation.
In this respect the smaller average contribution obtained in the SLy5
case makes it a better starting point. Nuclear matter properties can
also be used to refit, but then the nuclear matter EOS has to be calculated
at the corresponding level of many-body theory (see e.g. \cite{Moghrabi2012}
for a description of nuclear matter at MBPT2 order).

\section{Summary and conclusions}

The problem we set out to investigate was whether effective nuclear
interactions can provide improved descriptions of nuclear binding
energies when correlation effects beyond the HFB level are taken into
account. To this end, it was essential to introduce a momentum cut-off
in Skyrme's potential in order to obtain convergent results. The calculations
show that even with a low cut-off, the average part of the correlation
corrections are quite substantial (about 25 \% of the total binding
energy with MBPT2 and $\Lambda=1.8$ (fm)$^{-1}$). We then considered
a schematic renormalization by removing the average parts of the correlation
energies. The remaining fluctuations are similar for both interactions
studied and not so sensitive to the exact choice of the momentum truncation.
When the SLy5 Skyrme parametrization is used, the fluctuations associated
with low-lying surface vibrations does lead to a reduction of the
errors compared to experiment. In order to obtain a model that can
be used with more confidence a refit of the interaction parameters
should be performed. The ideal would be to compare results of an interaction
fitted on HFB level to those of an interaction fitted on the MBPT2
level using the same set of experimental data. 

Some interesting features can anyway be learned from the obtained
fluctuations. One result is that octupole vibrations are predicted
to give fluctuations of similar magnitude as the quadrupole corrections
and to contribute in a similar way. It is also interesting to see
that higher multipoles such as $4^{+}$ and $5^{-}$ gave rise to
large fluctuations in the case of Sn isotopes.

The fluctuating parts were extracted using MBPT2 but the QRPA formula
is also promising in the sense that it allows an infinite summation
of diagrams. However, in order for it to be a practical tool, the
QBA approximation must be improved and a careful study of the corrections
in the quasi-particle case would be needed. Once such a formalism
is in place, the correlation energy could be calculated using iterative
approaches \cite{Nguyen2009} similar in spirit to the ones we recently
employed for the calculation of low-lying excitations \cite{Carlsson2012}.

In summary, we have regularized Skyrme's potential and used it to
study higher order corrections to binding energies beyond the HFB
approach. Compared to other approaches, the method used here has the
advantage of not relying on energy truncations in order to converge
and that correlations resulting from many degrees of freedom (e.g.
vibrational modes) can be simultaneously included. Apart from nuclear
binding energies studied in this work, there are other quantities
that could possibly be modeled better in a formalism that goes beyond
the HFB approximation. An example is the calculation of alpha-decay
preformation amplitudes which shows a dramatic increase as correlations
between nucleons are introduced \cite{Lovas1998}. 

\begin{acknowledgments}
B.G. Carlsson acknowledges the Royal Physiographic Society in Lund for providing funding for the computers on which the calculations were performed. We also thank I. Ragnarsson for valuable comments on the manuscript. This work was supported in part by the Academy of Finland and University of Jyväskylä within the FIDIPRO programme.
\end{acknowledgments}

\bibliographystyle{apsrev4-1}
\bibliography{nuclear_references_v2}

\end{document}